\documentclass[prd,twocolumn,amsmath,amssymb,mathptmx,floatfix,nofootinbib,footnote,showkeys]{revtex4-2}
\usepackage{graphicx}
\usepackage{xcolor}
\usepackage{color}
\usepackage{bm}
\usepackage{float}
\usepackage{orcidlink}
\usepackage{url}
\begin{document}
\title [Large-scale vortex unpinning and pulsar glitches] 
{Large-scale unpinning and pulsar glitches due to the forced oscillation of vortices}
\author{Biswanath Layek\, \orcidlink{0000-0002-2544-2687}}
\email{layek@pilani.bits-pilani.ac.in}
\author{Brijesh Kumar Saini}
\email{p20230077@pilani.bits-pilani.ac.in}
\author{Deepthi Godaba Venkata\, \orcidlink{0000-0002-5240-5638}}
\email{p20210075@pilani.bits-pilani.ac.in}
\affiliation{Department of Physics, Birla Institute of Technology and Science, Pilani 333031, India}

\date{\today}
\begin{abstract}
The basic framework of the superfluid vortex model for pulsar glitches, though, is 
well accepted; there is a lack of consensus on the possible trigger mechanism 
responsible for the simultaneous release of a large number ($\sim 10^{17}$) of 
superfluid vortices from the inner crust. Here, we propose a simple trigger 
mechanism to explain such catastrophic events of vortex unpinning. We treat a 
superfluid vortex line as a classical massive straight string with well-defined 
string tension stretching along the rotation axis of pulsars. The crustquake-induced 
lattice vibration of the inner crust can act as a driving force for the transverse
oscillation of the string. Such forced oscillation near resonance causes the 
bending of the vortex lines, disturbing their equilibrium configuration and resulting 
in the unpinning of vortices. We consider unpinning from the inner crust's so-called 
{\it strong (nuclear)} pinning region, where the vortices are likely pinned to the nuclear sites.
We also comment on vortex unpinning from the interstitial pinning region of the inner crust. 
We sense that unifying crustquake with the superfluid vortex model can naturally explain 
the cause of large-scale vortex unpinning and generation of large-size pulsar glitches.
\end{abstract}

\maketitle
{\it Introduction :}
Although pulsars have an extraordinarily stable rotational frequency ($\Omega$), 
a significant of them are reported \footnote{\\http://www.jb.man.ac.uk/pulsar
/glitches/gTable.html} to show sudden spin-up events ({\it glitches}). The 
fractional change of rotational frequency, i.e., the 
size of glitches $\Delta \Omega/\Omega$, are observed to lie in the range $\sim 10^{-12} 
- 10^{-5}$ \citep{Espinoza_2011}, with the interglitch time varying from months to years. 
The first theoretical attempt to understand the physics behind such events, namely, 
the crustquake model, was proposed in the late 1960s \citep{1969Natur.223..597R, 1969Natur.224..872B}. 
Later, it was realized 
that the model suffers from compatibility issue \citep{baym1971}, as it demands a larger 
interglitch time to produce large-size glitches ($\Delta \Omega/\Omega \simeq 10^{-5} 
- 10^{-6}$) contrary to the observations. The subsequently proposed superfluid vortex 
model \citep{Anderson:1975zze}, which is widely accepted as a prime candidate among 
the pulsar glitch models \citep{Haskell:2015jra}, provides the proper framework for 
understanding such glitch events. However, even within the vortex model, the 
underlying mechanism of an instantaneous unpinning of an enormous number of vortices 
($\sim 10^{17}$) from the inner crust having a landscape of varying pinning energies 
is yet to be established. In this context, the role of crustquake either as a trigger 
mechanism for vortex-unpinning \citep{akbal2017} or as a source of thermal energy 
\citep{1996ApJ...457..844L, Layek:2020ocz, 
Layek:2022kja} has been discussed quite frequently in the literature. There were also suggestions 
that vortex avalanche \citep{PhysRevB.85.104503, Warszawski:2012wa} might be responsible for 
the required large-scale unpinning from the inner crust. Even with various suggestions, the 
resolution of the puzzle of large-scale vortex unpinning has yet to be settled.

Here, we suggest that by unifying crustquake with the superfluid-vortex model, one can 
naturally explain the simultaneous release of large-scale unpinning from the inner crust. 
We model the equilibrium configuration of a superfluid vortex line with a classical 
massive straight string with a tension $T$. We consider the 
{\it nuclear (strong) pinning region} \citep{Elgaroy:2001rg, Seveso_2015, Link_2022}, 
where each string passes through a number of pinning sites (heavy 
neutron-rich nuclei). We focus on a representative string segment with 
two ends pinned (fixed) to nuclear sites. The motion of a vortex line, supposedly 
a classical string, has been discussed earlier in Ref. \citep{Link_2022}.
The string parameter $T$, an essential parameter for determining the dynamics of 
vortex unpinning in the standard superfluid vortex model, also frequently appears in 
the literature \citep{Elgaroy:2001rg, Seveso_2015, Link_2022}. 

We assume a crustquake event triggers the inner crust lattice to vibrate in one of its normal modes.
The lattice vibration drives the string segments to execute force 
vibration and disturb the stability by bending the vortex lines. The pinned nuclear 
sites at two ends of a string segment provide the external driving forces. In 
a steady state, such a driving force can cause the string segment to resonate, thus 
maximizing the instability of the vortex lines, and as we will see, may result in 
the unpinning of superfluid vortices.

We implement our idea on the inner crust region, where the vortex-nuclear interaction 
has been suggested to be attractive. Hence, the vortex lines are preferably pinned to 
the nuclear sites, referred to as the nuclear pinning region in Ref. \citep{Seveso_2015}, 
or the strong pinning region in Ref. \citep{Link_2022}. Earlier, Epstein and Baym 
\citep{1988ApJ...328..680E} suggested that the nuclei bind the vortex lines for mass 
densities in the range $\rho \sim (10^{13} - 10^{14})~\mbox{g cm}^{-3}$ and are repelled 
at lower densities in the inner crust. However, the studies using the density function 
theory produces repulsive nucleus-vortex interaction upto $\rho \simeq  7 \times 
10^{13}~\mbox{g cm}^{-3}$ \citep{PhysRevLett.117.232701} (see also Ref. \citep{Link_2022}). 
The vortex lines in the weak pinning zone with less than above baryon density prefer 
interstitial pinning. Accordingly, we will consider the region $\rho \sim (7 \times 
10^{13} - 1.4 \times 10^{14})~\mbox{g cm}^{-3}$ (i.e., $ 0.3 \rho_0 - 0.6 \rho_0; 
\rho_0 = 2.4 \times 10^{14}~\mbox{g cm}^{-3}$) for the study of vortex unpinning by 
the forced vibration. The lattice vibration changes the dynamics of nuclear-vortex 
interaction and, hence, should affect the vortex lines even in the interstitial 
pinning region (weak pinning region). However, we will not study this case here. 
Instead, we assume the standard {\it knock-on} process for vortex unpinning for this 
region. In the so-called knock-on process, the unpinned vortices in the strong pinning 
zone, while moving outward, can knock on and unpin the vortices from the interstitial 
pinning region. The knock-on process has been discussed in detail in the literature \citep{PhysRevB.85.104503, 
Warszawski:2012wa} (see also \citep{Layek:2022kja} for a semi-quantitative implementation 
of this process.). Finally, to realize the large-scale release of inner-crust 
vortices, our proposed picture of vortex unpinning from the strong pinning region should 
be supplemented with the knock-on process (for the weak pinning zone). Below, we first 
briefly review the necessary ingredients of our model and then present the numerical 
estimation of various relevant quantities, followed by results \& discussion. Finally, 
we conclude with some comments on further scopes of studies.\\

\begin{figure}
\centering
\includegraphics[width=1.0\linewidth]{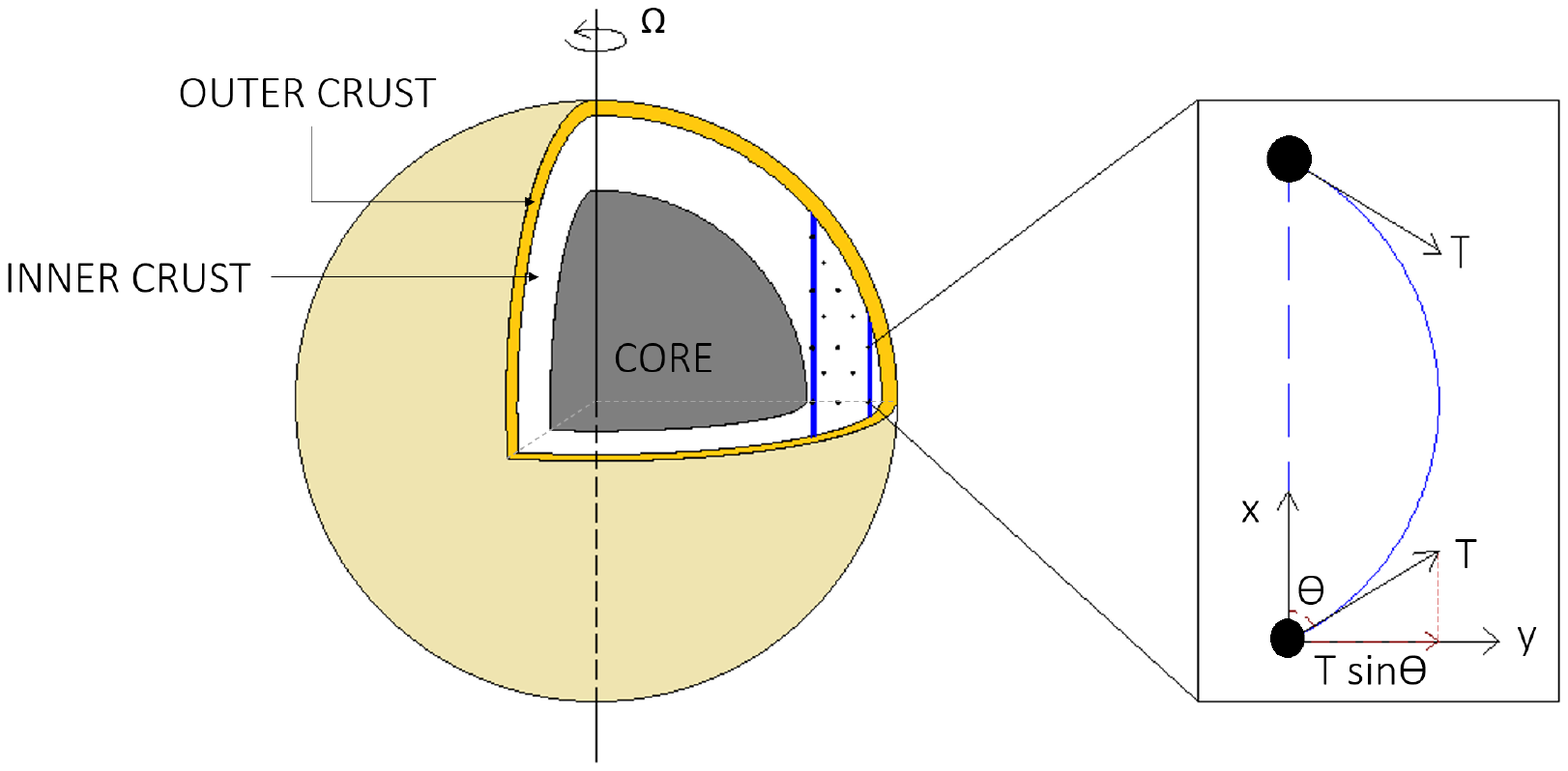}
\vspace{-0.5in}
\caption{A schematic picture of vortex lines (blue) in the inner crust of a neutron 
star rotating with angular frequency $\Omega$. Each vortex line passes through several 
nuclear sites (black dots). On the right, a representative string segment undergoing 
forced oscillation driven by lattice vibration is enlarged for clarity (the picture is not 
to scale).}
\label{fig:fig1}
\end{figure}

{\it Superfluid Vortex Vibration :} 
We take a representative string segment of length $l$, tension $T$, and mass per 
unit length $\mu$. The string segment is assumed to extend along the axis of rotation 
and oscillate in the transverse $x$-$y$ plane as shown in Fig. \ref{fig:fig1}.
Two ends of the string segment are pinned (fixed) to nuclear sites. The wave equation 
representing the string vibration can be written as  
\begin{equation}
\frac{\partial^2 y (x,t)}{\partial t^2} = v^2 \frac{\partial^2 y (x,t)}{\partial x^2} 
- \gamma \frac{\partial y (x,t)}{\partial t}.
\label{eq:waveeq}
\end{equation}
Where $v = \sqrt{T/\mu}$ is the velocity of the wave propagating along the string, and the term 
involving $\gamma$ takes care of the damping effect. Let us first recall the allowed set 
of normal modes $\{\omega_n;  n=1,2,...\}$ with fixed boundaries $y(0,t) = y(l,t) = 0$. 
The frequencies of the fundamental mode $\omega_1$ and the $n$th mode $\omega_n$ are 
given by \citep{french}
\begin{equation}
\omega_1 = \pi v/l ~;~ {\omega_n = n \omega_1}.
\label{eq:omegas}
\end{equation}
We will estimate later the numerical values of ${\omega_n}$ by taking suitable 
values of various parameters associated with the string segment.\\
  
{\it Inner Crust Lattice Vibration :}
We assume a one-dimensional linear chain consisting of nuclear sites with equilibrium 
inter-nuclear distance $a$. Considering the nearest neighbor interaction and harmonic 
potential approximation, the dispersion relation of the wave in the lattice can be written 
as \citep{kittel}
\begin{equation}\omega = \omega_{max} \sin (\frac{ka}{2});~ 
\omega_{max} = \sqrt{\frac{4\alpha}{m}}.
\label{eq:omegam}
\end{equation}
Where $\omega_{max}$ is the maximum value of the normal mode frequencies and $\alpha$ is the effective
spring constant. In the long wavelength limit $k \rightarrow 0$, the group velocity $v_g$ and 
the phase velocity $v_p$ are the same and related to the sound's speed in the inner crust material, 
i.e.,  $v_s = v_g = (\omega_{max} a)/2$. Using periodic boundary condition, the frequency difference 
between successive normal modes is given as $\Delta \omega = v_s \Delta k = 2 \pi v_s/L$ 
; $L$ is the lattice size, taken to be the thickness of the inner crust ($\sim 1$ km). 
Since the inter-nuclear distance in the inner crust $a$ is in the femtometer range, the frequency 
deference $\Delta \omega = \pi (a/L) \omega_{max}$ is very small relative to $\omega_{max}$. Thus, 
the lattice's normal mode frequency can be assumed to vary continuously from $0$ to 
$\omega_{max}$. \\

{\it Forced Vibration and Shape of a Vortex Line :}
Our basic picture is the forced vibration of vortex lines driven by the
oscillatory nuclear sites to which the vortex lines are pinned.
The external driving force can be implemented through suitable boundary conditions 
(BC) $y(0,t) = d_0 \cos (\omega t)$ and $y(l,t) = d_l \cos (\omega t)$. Here, we assume 
the oscillation of the nuclear sites with a normal mode frequency $\omega$ with amplitudes 
$d_0$ at $x = 0$ and $d_l$ at $x = l$. First, we consider 
the steady state solution $y(x,t) = f(x) \cos (\omega t)$ of Eq.(\ref{eq:waveeq}) for
undamped motion ($\gamma = 0$). The solution can be written in terms of the wave 
vector $k (= \omega/v)$ as
\begin{equation}
y (x,t) = d_0 \left[\frac{\cos [k(x-l/2)]}{\cos (kl/2)} 
+ (r-1) \frac{\sin kx}{\sin kl} \right] \cos (\omega t) .
\label{eq:sol1}
\end{equation}
Where $r= d_l/d_0$ is the ratio of the amplitudes of two oscillatory nuclear sites 
attached across a few femtometer string segments and the ratio is expected to be 
approximately one. Since $\cos (kl/2) \simeq \cos [(\pi/2) (\omega/\omega_1)]$, 
for $r=1$, the string amplitude $f(x)$ is clearly enhanced when the driving frequency 
$\omega \simeq (2n+1)~\omega_1$. If $r$ deviates significantly 
from unity, the amplitude enhancement occurs for $\omega$ close to any of the 
string's normal mode frequencies $\{\omega_n;  n=1,2,...\}$.
Note, $f(x)$ diverges at resonance, and the solution is invalid for $\omega = \omega_n$. 
Now, we consider the realistic vortex motion in the presence of vortex drag force 
($\gamma \ne 0$). For such a case, the wave propagating along increasing $x$ takes 
the form $y(x,t) \propto e^{i(Kx - \omega t)}$; with the wave vector should be treated
complex $K = k_1 + i k_2$. Where,
\begin{equation}
k_1 \simeq k (1 + \frac{\gamma^2}{8\omega^2}) = \frac {\omega}{v} 
(1 + \frac{\gamma^2}{8\omega^2});~ k_2 \simeq \frac{\gamma}{2v}.
\label{eq:k1}
\end{equation}
The real part of the wave vector $k_1$ is responsible for the wave propagation, while 
$k_2$ is for the wave's attenuation. The solution 
of Eq.(\ref{eq:waveeq}) with the same BCs as earlier $y(0,t) = d_0 \cos (\omega t)$ 
and $y(l,t) = d_l \cos (\omega t)$, can be written by replacing $k$ with 
$k_1 \simeq k (1 + \gamma^2/8\omega^2)$ in Eq.(\ref{eq:sol1}),
\begin{equation}
y (x,t) = d_0 \left[\frac{\cos [k_1(x-l/2)]}{\cos (k_1l/2)} 
+ (r-1) \frac{\sin k_1x}{\sin k_1l} \right] \cos (\omega t).
\label{eq:sol2}
\end{equation}
Where,
\begin{equation}
k_1 l \simeq kl + \frac{\gamma^2 l}{8v\omega} = 
\frac{\pi \omega}{\omega_1} + (\frac{\pi}{8}) 
(\frac{\omega}{\omega_1}) (\frac{\gamma}{\omega})^2. 
\label{eq:reson}
\end{equation}

We have ignored the attenuation factor $e^{-k_2 x}$ in Eq.(\ref{eq:sol2}) due
to the following reason. 
The attenuation length $l_a$ is determined by the propagation distance from the source, 
where the wave amplitude decays to $1/e$ and is given by $l_a = 1/k_2 = 2v/\gamma$. 
Thus, $e^{-k_2 x}$ term can be ignored, provided $l_a = 1/k_2 = 2 v/\gamma >> l$, 
i.e., $\gamma/\omega_1 << 0.6$. This condition is satisfied as 
$\gamma/\omega_1 \sim 10^{-2}$ (shown later). \\

Now, using Eq.(\ref{eq:sol2}), the amplitude $f(x)$ at resonance can be written as (for $r=1$)
\begin{equation}
f(x) \simeq (-1)^{n+1} d_0 \frac{16(2n+1)}{\pi} (\frac{\omega_1}{\gamma})^2 
\cos \left[k_1(x-l/2)\right].
\label{eq:sol3}
\end{equation}
Here, we have used $\cos [(2n+1)\frac{\pi}{2} + \epsilon] \simeq (-1)^{n+1} \epsilon$
($n = 0,1,2,...$) for small $\epsilon = \frac{\pi}{16(2n+1)} (\frac{\gamma}{\omega_1})^2$. The unpinning of a 
vortex line from the pinning site due to the deviation from its equilibrium configuration 
can be best understood by calculating the force on the nuclear site acting radially 
(see, Fig. \ref{fig:fig1})
$F_y = T \sin \theta = T f^\prime(x) /\sqrt{1 + {f^\prime(x)}^2}$ at $x = 0$ 
(or, equivalently at $x=l$). The force $F_y$ is a crucial quantity that effectively determines 
the unpinning force and needs to be compared with the pinning force $F_p$ per nuclear site. 
The slope at $x = 0$ can be calculated using Eq.(\ref{eq:sol2}) as
\begin{equation}
f^\prime(0) = d_0 k_1 \left[\tan (k_1 l/2) + \frac{(r-1)}{\sin k_1l}\right],  
\label{eq:slope}
\end{equation}
which is given as $f^\prime(0) = d_0 k_1 \frac{16(2n+1)}{\pi} 
(\frac{\omega_1}{\gamma})^2$ (for $r=1$) at resonance $\omega = (2n+1) \omega_1$. 
Thus, the unpinning force $F_y = T \sin \theta = T f^\prime(0) 
/\sqrt{1 + {f^\prime(0)}^2} \simeq T$ is significantly enhanced 
near resonance, providing enough scope for vortex-unpinning. \\

{\it Numeric :} 
To determine the string's fundamental frequency $\omega_1$, we must first 
calculate the wave propagation velocity $v~(= \sqrt{T/\mu})$ along the 
string. One can calculate $v$ by taking a suitable value of linear mass 
density of the vortex $\mu$, and the vortex tension $T$ 
\citep{PhysRevLett.102.131101, Link_2022}. For the estimate of $\mu$, 
we assume the vortex as a cylinder with a sharp radius filled with
uniform normal neutrons of density $n$ (see Ref. \citep{Elgaroy:2001rg} 
for such an assumption and its validity). Taking the coherence length  $\xi$ of 
neutron-neutron cooper pairs as an approximate size of the vortex core ($\sim$ 20 fm), 
one obtains $\mu = \pi \xi^2 m_n n$. As vortex tension is one of the most important 
quantities (pinning energy is another) for understanding the dynamics of vortex 
unpinning, the numerical estimate of $T$ is paramount. We use the recent result, 
where a hydrodynamic estimate of the vortex tension is written as \citep{Link_2022}
\begin{equation}
T = 0.6 \left(\frac{\rho_s}{10^{13}~\mbox{g cm}^{-3}}\right)~\mbox{MeV/fm}.
\label{eq:tmu}
\end{equation}
Where $\rho_s$ is the (unentrained) superfluid mass density, roughly the same as the stellar 
mass density $\rho$ for the strong pinning region \citep{1992ApJ...387..276E}. For $\rho = 
10^{14}~\mbox{g cm}^{-3}$, the value of $T = 6 ~\mbox{MeV/fm}$ is somewhat smaller than the 
values provided in Refs. \citep{PhysRevLett.117.232701,Elgaroy:2001rg}. However, since in 
our case, the ratio $F_y/F_p$ determines the vortex dynamics, we will assume the result of 
Eq.(\ref{eq:tmu}) and take the vortex-nucleus pinning force $F_p$, or equivalently the 
pinning energy $E_p$, to be consistent with the above string tension. From the knowledge 
of the pinning energy, one can estimate the pinning force $F_p \simeq E_p/\xi$. The pinning 
energy varies across the inner crust and lies in the 
range (0.72 - 0.02) MeV \cite{Seveso_2015} for $\rho \simeq (7 \times 10^{13} - 1.4 \times 
10^{14})~\mbox{g cm}^{-3}$. Thus, the pinning force can be taken as $F_p \simeq (3.6 
\times 10^{-2} - 10^{-3}$) MeV/fm. \\

Now, the propagation velocity $v$ can be estimated as
\begin{equation}
\frac{v}{c} \simeq 2 \times 10^{-3}\left(\frac{\rho_s}{10^{13} ~\mbox{g cm}^{-3}}\right)^{1/2} 
\left(\frac {n_0}{n}\right)^{1/2}\left(\frac{20~\mbox{fm}}{\xi}\right).
\label{eq:vbyc}
\end{equation}
Where $n_0$ ($\sim 0.16~\mbox{fm}^{-3}$) and $c$ are the nucleon saturation density and 
the speed of light in vacuum, respectively. Numerically, the string's fundamental 
frequency $\omega_1$ can be determined from Eq.(\ref{eq:omegas}) and Eq.(\ref{eq:tmu}) 
as
\begin{eqnarray}
\label{eq:omega1} 
\omega_1 &\simeq&  1.7 \times 10^{18} \left(\frac{\rho_s}{10^{13} ~\mbox{g cm}^{-3}}\right)^{1/2} 
\left(\frac {n_0}{n}\right)^{1/2} \nonumber \\
 &\times&  \left(\frac{20 ~\mbox{fm}}{\xi}\right)
\left(\frac{10^3~\mbox{fm}}{l}\right)  s^{-1}. \\
\nonumber
\end{eqnarray}
Where $n = (0.3 - 0.6) n_0$ is the baryon density for the relevant part of the inner crust 
region. We take the length of each string segment $l \simeq 10^3$ fm and an average inter-nuclear 
distance $a \simeq 50$ fm. The above choice of $l$ is consistent with the fact that for the 
vortex with finite tension, the ratio $l/a$ should be much larger than unity \citep{Link_2012} 
for successfully bending and pinning. Also, as noted in Ref. \citep{Link_2012}, the ratio 
$l/a \sim 32$ for the pinning energy $E_p = 0.1$ MeV. Finally, for $\rho$ in the range  
$(7 \times 10^{13} - 1.4 \times 10^{14})~\mbox{g cm}^{-3}$ ($n = 0.3 n_0 - 0.6 n_0$), 
the fundamental frequency for a $10^3$ fm string segment is given by $\omega_1 \simeq 8.2 
\times 10^{18}~\mbox{s}^{-1}$.

As far as the normal mode frequency of the lattice is concerned, the maximum value
$\omega_{max}$ (see Eq.(\ref{eq:omegam})) is set by the sound's speed $v_s$ and the inter-nuclear 
distance $a$. Using the appropriate value of $v_s \simeq 0.2 c$ for 
the inner crust materials \citep{Ecker_2022} and $a \simeq 50$ fm, one obtains 
$\omega_{max}  = 2v_s/a \simeq 2.4 \times 10^{21}~\mbox{s}^{-1}$. Finally, the value of 
damping coefficient $\gamma$ appearing in Eq.(\ref{eq:sol2}) 
is estimated in Ref. \citep{Link_2022} (see also  \citep{1992ApJ...387..276E}), 
where the drag force (per unit length) $f_l$ on the vortex 
line moving with velocity $\dot y$ is expressed as $f_l = - \eta \dot y$ with 
$\eta/\rho_s \kappa \sim 10^{-3}$. Comparing with our definition $f_d = \gamma \mu \dot y$
in Eq.(\ref{eq:waveeq}), we obtain $\gamma = \eta/\mu \simeq 10^{17}~\mbox{s}^{-1}$,
i.e., $\gamma/\omega_1 \sim 10^{-2}$. Where $\kappa = h/2m_n$ is the magnitude of the 
vorticity associated with the neutron superfluid, and we have taken 
$\rho_s \sim 10^{14}~\mbox{g cm}^{-3}$.\\

{\it Results and Discussions :}
For an effective vortex unpinning, the driving frequency $\omega$ should lie 
in the range $\omega_1 (\sim 10^{19}~\mbox{s}^{-1})$ to $\omega_{max} (\sim 10^{21}~\mbox{s}^{-1})$. 
The possibility of unpinning is pronounced when $\omega \simeq  n \omega_1$. Here, we shall provide 
a rough estimate of the frequency at which the lattice is most likely to vibrate. For this, we take 
the standard picture of crustquake \citep{baym1971} and assume that strain energy released ($\Delta E \sim 10^{40}$ erg)  
in a crustquake triggers the inner crust containing $N_0 \sim (L/a)^3 \sim 10^{46}$ nuclear 
sites to vibrate with frequency $\omega_0$. Equating $N_0 \hbar \omega_0$ with the strain energy 
provides $\omega_0 \simeq 10^{21}~\mbox{s}^{-1}$. Interestingly, the estimated value is close to the inner 
crust lattice's maximum allowed normal mode frequency, $\omega_{max}$. We now roughly estimate the amplitude 
$d_0 (\simeq d_l)$ of the oscillatory pinning site of mass $M$ consisting of $N_n$ nucleons. This can be 
estimated by equating $(1/2) M \omega_0^2 d_0^2$ with $\hbar \omega_0$. Taking the number of nucleons 
in a neutron rich nucleus in the range, $N_n = (180 - 1800)$ \citep{Pastore:2011qk}, the values of 
$d_0 = 14/\sqrt{N_n}$ fm are observed to lie in the range (1.0 - 0.3) fm. \\
\begin{table}
\centering
\caption{Fiducial values of parameters.}
\label{tab:table}
\begin{tabular}{lccc}
\hline
Quantity &~~~& Value\\
\hline
Mass density ($\rho$) &~~~& $10^{14}~\mbox{g cm}^{-3}$ \\
Lattice spacing ($a$) &~~~ &  50 fm      \\
Coherence length ($\xi$) &~~~ & 20 fm \\
Vortex tension ($T$) &~~~ & $6~\mbox{MeV fm}^{-1}$ \\
Length of string segment ($l$) &~~~ & $10^3$ fm \\
Pinning force ($F_p$) &~~~ & $10^{-3} ~\mbox{MeV fm}^{-1} $\\
\hline
\end{tabular}
\end{table}
\begin{figure}
\centering
\includegraphics[width=1.0\linewidth]{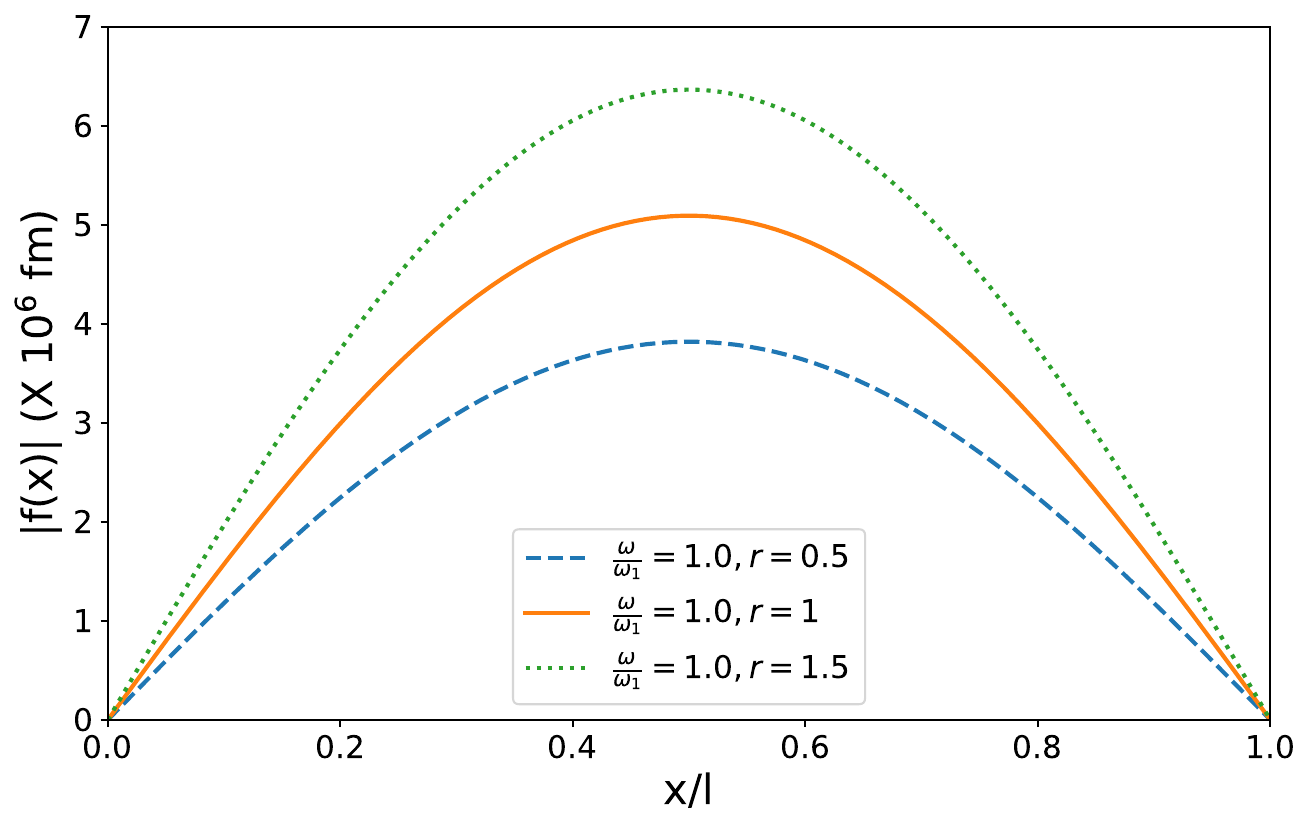}
\includegraphics[width=1.0\linewidth]{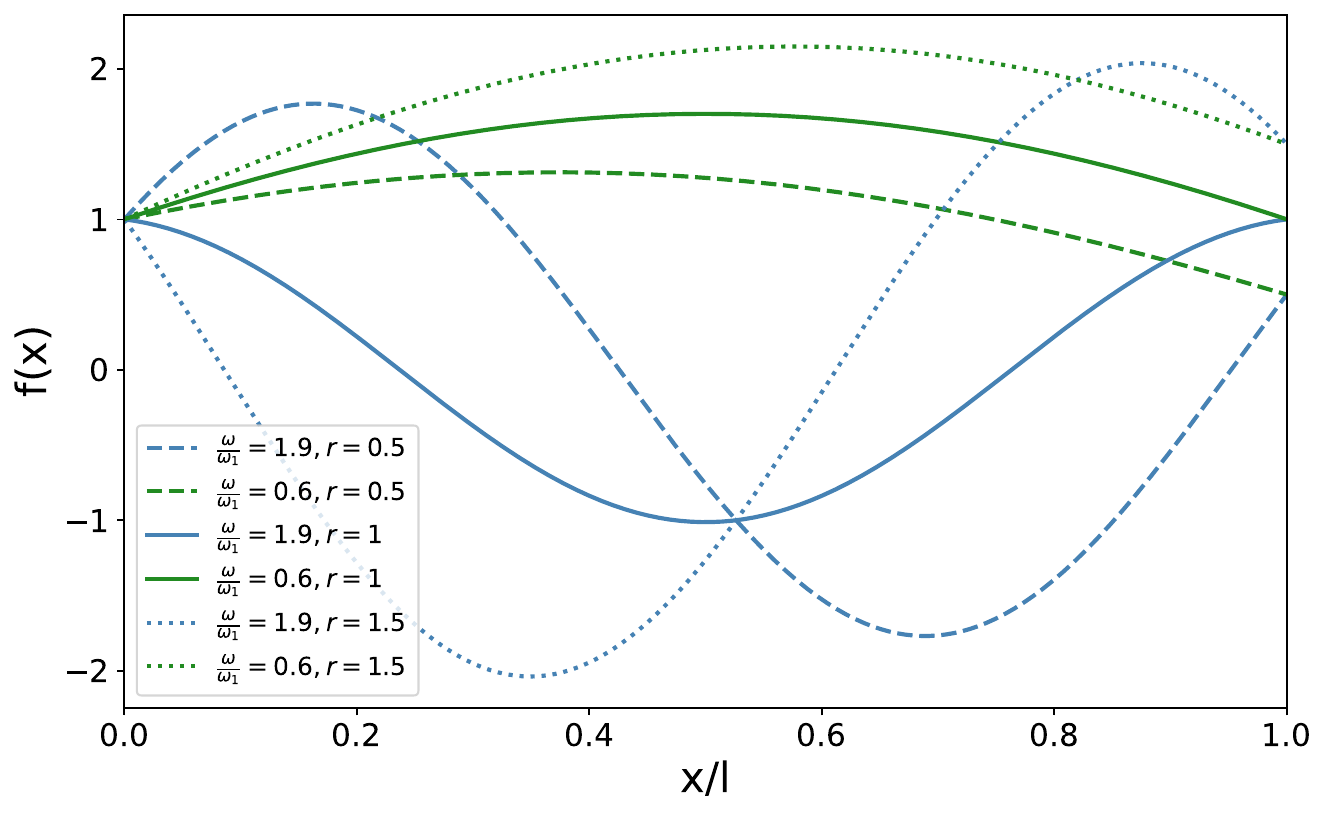}
\includegraphics[width=1.0\linewidth]{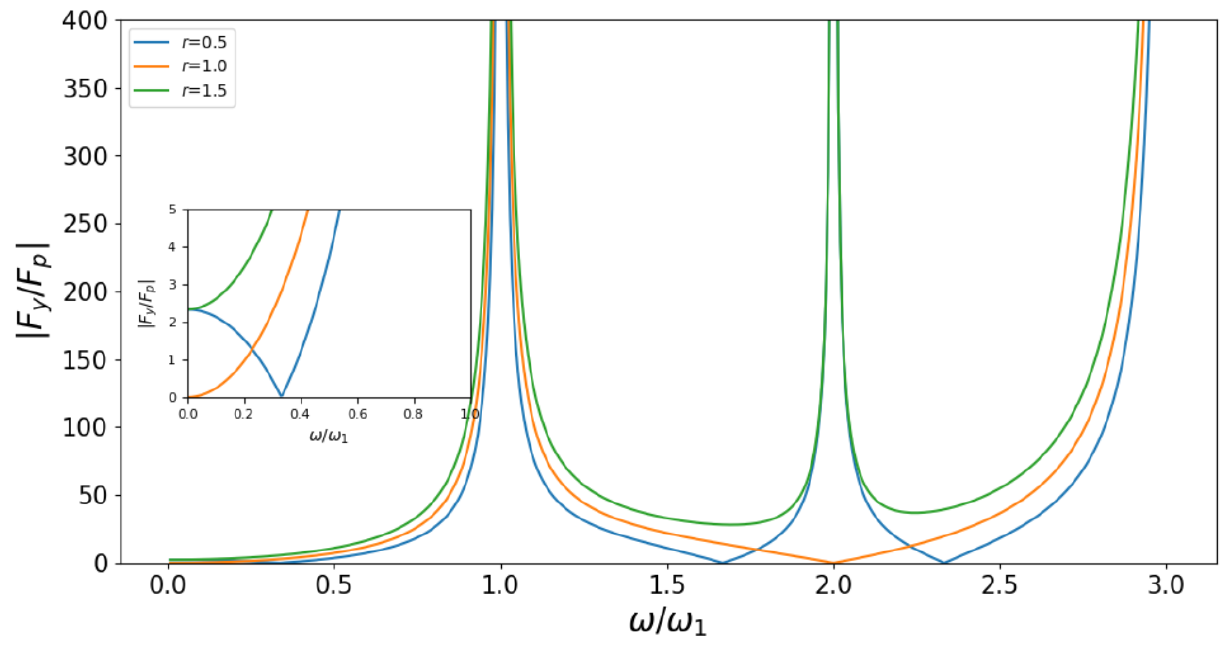}
\caption{(Top) String amplitude $f(x)$ at various locations of the string segment at resonance 
( i.e., at $\omega = \omega_1$) by taking sample values of $r$ ($= d_l/d_0$) = 1, 1.5 and 0.5. 
(Middle) String amplitude at a few arbitrarily chosen driving frequencies $\omega$ around the 
string's fundamental frequency $\omega_1$. (Bottom) The variation in the ratio of unpinning force 
$F_y (= T \sin \theta)$ to the pinning force $F_p$ for a range of driving frequencies $\omega$. 
See Table \ref{tab:table} for the values of various parameters.}
\label{fig:fig2}
\end{figure}

Here, we present our analysis by taking $d_0 = 1$ fm. Also, we will only show the results 
for the driving frequencies around $\omega_1$, as a similar effect is expected for other 
values of $\omega = n \omega_1$.
 Fig. \ref{fig:fig2} (top plot) shows the maximum variation 
of the string's configuration at resonance ($\omega = \omega_1$) for arbitrarily chosen 
a few values of the ratio $r = d_l/d_0$. Fig. \ref{fig:fig2} (middle plot) shows the 
same, but away from resonance. Again, the disturbance of the string's equilibrium position 
is visible, though small, as expected, compared to the case at resonance. The unpinning force
$F_y (= T \sin \theta)$ arises due to the deviation of the string's equilibrium configuration 
is shown in Fig. \ref{fig:fig2} (bottom plot), where, for comparison, the ratio $|F_y/F_p|$ 
(keeping $F_p$ fixed) is plotted versus the driving frequency $\omega$. Clearly, for
$F_p \simeq 10^{-3}$ MeV/fm and $r=1$, there is an appreciable deviation of the string's equilibrium 
configuration in the frequency range $\omega = (0.2 - 1.9)~\omega_1$, providing enough 
unpinning force ($|F_y/F_p| \ge 1$) on the nuclear site. 
For $r \sim 1.5$, the unpinning seems to occur for all values of 
$\omega$, while for $r \sim 0.5$, there are a few gaps in $\omega$ for which unpinning might not
occur. Note, for $r \ne 1$, the second term in Eq.(\ref{eq:sol2}), though small compared 
to the first term, can have a significant contribution, particularly at 
$\omega = n \omega_1 (n = 2, 4,...)$. Including such a term allows both normal mode frequencies, 
even and odd, to produce the resonance phenomenon. However, for a string segment of length 
$l \simeq 10^3$ fm, the results for $r \simeq 1$ (i.e., the approximately equal amplitude of 
vibration of the nuclear sites pinned at two ends of a line segment) may be more reliable.  
 
Here are a few observations : (1) As the driving frequency's precise value is unknown, we provide 
a possible range $\Delta \omega \sim \omega_1$ around $\omega_1$ for which the 
vortex unpinning is expected to be effective. (2) The effect is not specific to around $\omega_1$ only. 
A similar phenomenon should be observed for $\omega$ closed to any other normal mode frequencies 
$\{\omega_n (\le \omega_{max})\}$ with a similar range $\Delta \omega \sim \omega_n$. 
The number of such normal modes within $\omega_1$ to $\omega_{max}$ is also reasonably large 
$\omega_{max}/\omega_1 \sim 100$. Thus,  the resonance phenomenon is likely effective for arbitrary 
driving frequency $\omega~(\le \omega_{max})$.
(3) The results are presented by taking a sample value
$\omega_1 = 8.2 \times 10^{18}~\mbox{s}^{-1}$. However, the numerical value of 
$\omega_1 = (\pi/l) \sqrt{T/\mu}$ depends on the parameters associated with the 
vortex lines, such as string tension $T$, length of a string segment $l$, and 
mass per unit length $\mu$ of a vortex line. Since these parameters vary across 
the inner crust (see Eq. (\ref{eq:omegas}) and Eq. (\ref{eq:tmu})), in principle, 
the specific value of $\omega_1$ should represent a particular region of the 
inner crust. However, it is evident from Eq. (\ref{eq:omega1}) that for a fixed value of 
$l$, numerical value of $\omega_1$ is almost constant throughout the strong pinning region. 
Thus, $\omega_1$ mainly depends 
on the length of the string segment. For the lattice spacing of about 50 fm, 
we have considered a reasonable string length $l = 10^3$ fm, as suggested in the literature. 
Order of magnitude shorter $l$ (e.g., $l \simeq 2 a$) produces an unrealistic pinning 
force \citep{Link_2012}. On the contrary, larger $l$ reduces $\omega_1$ and increases 
the ratio $\omega_{max}/\omega_1$. Thus, larger $l$ adds more normal modes of a string 
segment and increases the possibility of frequency matching ($\omega$ with $\omega_n$). 
All the above points coherently ensure that the resonance is inevitable everywhere in 
the strong pinning zone, irrespective of the values of the driving frequency 
$\omega$ ($\omega_1 \lesssim \omega \le \omega_{max}$).

The number of affected vortices due to the resonance in the strong pinning region 
can be written as $N_v = 2 \pi R_{in} D_s n_v$. Where $D_s$ is the thickness 
of the region, $R_{in}$ ($\sim 9$ km) is the inner crust radius, and $n_v = 4m_n 
\Omega/h \simeq 10^7 \text{m}^{-2}  (\Omega /\text{s}^{-1})$ is the areal vortex density. 
The precise thickness of the strong pinning region depends on a specific neutron star structure 
model. For a typical $1.4 \mbox{M}_\odot$ neutron star with an equation of state of moderate 
stiffness, the value of $D_s$ turns out to be about 400 m 
(see Ref. \citep{1996ApJ...457..844L} and the references therein). Thus, the number of unpinned vortices, 
say, for Vela pulsar with $\Omega \simeq 70~\mbox{rad/s}$,  is given by $N_v \simeq 10^{16}$.
These unpinned vortices, while moving outward can {\it knock-on} 
\citep{PhysRevB.85.104503, Warszawski:2012wa} and 
unpin more vortices \citep{Layek:2022kja} from the weak pinning zone, providing the 
possible explanation of large-scale (almost) instantaneous vortex unpinning. 
The crustquake which releases strain energy and triggers lattice vibration is assumed to occur 
frequently, with an approximately one-year waiting period \citep{1971AnPhy..66..816B} 
between any two successive glitches (set by a typical frequency of Crab-pulsar glitches). 
Such waiting period is crucial enough for the pinned superfluid-vortex region 
to carry an extra angular momentum $\Delta \Omega$ compared to the rigid (outer) 
crust-core region. The sudden release of these vortices caused by the forced vibration,
followed by a knock-on process providing the necessary large-size glitches.\\

{\it Comments and Scopes:}
Our proposal aims to understand the physics behind the large-scale vortex unpinning 
from the inner crust. We observe that unifying the crustquake and superfluid-vortex 
models can provide a natural explanation of such events. Further, crustquakes with 
about one event per year can explain the observed large-size glitches without any 
contradiction of producing large-size glitches with a small waiting period. 
This is because the crustquake here merely triggers, while the instantaneous release 
of vortices produces glitches.  

We initiated the proposed study by taking a sample of a string segment representing a 
specific region of the inner crust. There is a scope for extending the study by taking 
a single vortex line consisting of several string segments lying within the strong 
pinning region but with varying string and lattice parameters suitable to the specific 
areas (Note, a vortex line extending along $x$-axis crosses through a varying baryon 
density region.). One can also simulate the entire strong pinning zone to observe the 
overall effect. For such a study, identifying the strong pinning region and determining 
its thickness can be achieved by constructing a standard neutron star structure with a 
suitable equation of state. Similarly, one can extend the study for the interstitial 
pinning region to see if the lattice vibration also affects the vortex pinning in that 
region. Other possible consequences of the oscillation of vortex lines, such 
as the heating effect due to the excitation of Kelvin waves \citep{1992ApJ...387..276E} 
can be explored to constrain various parameters, particularly the self-energy of vortex 
lines, the length between any two successive pinning sites, pinning energy, etc. 
We want to explore some of these in our subsequent work. \\

{\it Acknowledgments :}
We thank Arpan Das for his valuable suggestions and Rashmi R. Mishra 
for helpful discussions. \\

{\it Data Availability :}
No new data were generated or analyzed in support of this research.

\bibliography{stringy} 
\end{document}